# Electrical Transport Properties of Graphene Nanoribbons Produced from Sonicating Graphite in Solution

Cheng Ling[1,*], Gabriel Setzler[1,2,*], Ming-Wei Lin[1,*], Kulwinder , Jin Jin[1], , Hyeun Joong Yoon[2], Seung Soo Kim[1,2], Mark Ming-Cheng Cheng[2], Noppi Widjaja[1] and Zhixian Zhou[1, a)]

[1]Department of Physics and Astronomy, Wayne State University,
Detroit, MI 48201

[2]Department of Electrical and Computer Engineering, Wayne State University,
 Detroit, MI 48202

Abstract

A simple one-stage solution-based method was developed to produce graphene nanoribbons by sonicating graphite powder in organic solutions with polymer surfactant. The graphene nanoribbons were deposited on silicon substrate, and characterized by Raman spectroscopy and atomic force microscopy. Single-layer and few-layer graphene nanoribbons with a width ranging from sub-10 nm to tens of nm and length ranging from hundreds of nm to 1 μm were routinely observed. Electrical transport properties of individual graphene nanoribbons were measured in both the back-gate and polymer-electrolyte top-gate configurations. The mobility of the graphene nanoribbons was found to be over an order of magnitude higher when measured in the latter than in the former configuration (without the polymer electrolyte), which can be attributed to the screening of the charged impurities by the counter-ions in the polymer electrolyte.   This finding suggests that the charge transport in these solution-produced graphene nanoribbons is largely limited by charged impurity scattering.

* These authors contributed equally.

a) Author to whom correspondence should be addressed, electronic mail: zxzhou@wayne.edu

**Introduction**

Graphene is an interesting material for both fundamental research and practical applications[1]. Several different methods have been developed to produce graphene, including mechanical cleavage [2], chemical vapor deposition, epitaxial growth [3], and exfoliation of graphite (as well as its derivatives and intercalation compounds) in solution [4, 5]. The solution method is not only scalable but also well-suited for chemical functionalization, creating opportunities for a wide range of applications [6]. Graphene oxide can be easily exfoliated in solution and subsequently reduced to graphene. However, the harsh oxidation process used to synthesize graphite oxide leaves functional groups on graphene; and removing such oxygen containing groups through chemical or thermal reduction simultaneously produces structural defects [7-9], which unavoidably degrade the electrical properties of the graphene. Coleman's group has recently demonstrated that large quantity of high-quality unoxidized graphene could be produced and dispersed in various organic solvents [5, 10, 11]. However, the electrical transport properties of graphene produced from the exfoliation of unoxidized graphite in solution remain largely unexplored, in spite of their relevance to various electronic applications. A reported transport study of individual few-layer graphene produced by exfoliating graphite in organic solvent showed no gate-voltage dependence of the conductance, while gate tunability is an essential characteristic of high quality graphene [12]. The lack of gate tunability in these dielectrophoretically assembled individual graphene devices was attributed to the relatively small gate voltage window [12]. Therefore, further electrical transport studies of solution-produced graphene are necessary for understanding their charge transport mechanism and exploring their potential for various electronic applications.

In this article, we demonstrate a simple one-stage synthesis method to produce graphene nanoribbons (GNRs) with large length-to-width ratios and straight edges by sonicating graphite powder directly in organic solution, without any prior chemical treatment of the graphite. While graphene is a zero gap semiconductor with finite minimum conductivity, which poses a major problem for conventional digital logic applications, a band gap opens up in narrow GNRs through spatial confinement and edge-effects [13]. To understand the transport mechanism and shed light on the nature of the remaining disorder in our GNRs, we fabricated field effect transistor (FET) devices consisting of individual few-layer GNRs and measured their transport properties in both the back-gate and polymer-electrolyte top-gate configurations. An order of magnitude mobility increase is observed in the latter configuration than in the former configuration (without the polymer electrolyte) due to the ionic-screening effect of the polymer electrolyte, suggesting that the charge transport in these GNRs is largely limited by charged impurity scattering.

**Experimental details**

Sieved graphite powder (10mg, Sigma Aldrich) was dispersed in a solution of 10 ml 1,2-dichloroethane (DCE) and 2 mg poly(m-phenylenevinylene-co 2, 5-diy octocy- p-phenylenevinglene) (PmPV) by bath sonication (Branson 3510 ultrasonic cleaner) for 1 hr. After sonication, we obtained a homogeneous black suspension of graphene sheets and a large amount of macroscopic aggregates. The dispersion was then briefly centrifuged (5 minutes) at 15000 rpm (Fisher Scientific Marathon 26kmr centrifuge) to remove aggregates and larger graphene sheets; and a supernatant containing thin graphene sheets and nanoribbons was obtained.

Raman spectroscopy and non-contact mode atomic force microscopy (AFM, Park System XE-70) were used to characterize the graphene samples deposited on Si/SiO$_2$ substrates from the supernatant. Raman Spectra were collected using a Jobin–Yvon Horiba Triax 550 spectrometer, a liquid-nitrogen cooled charge-coupled device (CCD) detector, an Olympus model BX41 microscope with a 100 × objective, and a Modu-Laser (Stellar-Pro-L) Argon-ion laser operating at 514.5 nm. The laser spot size was a few micros and the laser power at the sample was maintained at low level (~ 2 mW) to avoid any heating effect. To determine the ribbon width, the artifical width increment ($\Delta W$) due to tip dilation as determined by both the ribbon thickness ($h$) and tip radius ($R$): $\Delta W = 2[h(2R-h)]^{1/2}$ is subtracted from the apparent widths in the AFM images using the measured ribbon thickness and estimated tip radius provided by the tip manufacturer [14].

To characterize the electrical transport properties of GNRs, we fabricated FET devices of individual GNRs deposited on degenerately doped Si substrates with 285 nm of thermal $SiO_2$, where the Si substrate was used as the back gate. To remove the PmPV and solvent residue, the substrates were heated in air at 375 °C for 15 minutes and annealed at 600 °C for 10 minutes with flow of forming gas to further clean the samples and substrate [15]. Subsequently, non-contact mode AFM was used to locate and characterize GNRs with respect to prefabricated Au alignment marks. Electrodes were fabricated on selected GNRs using standard electron beam lithography followed by electron beam assisted deposition of a 1 nm Cr adhesion layer and 50 nm of Au. The devices are annealed in vacuum at 600 °C for 10 minutes to clean the ribbons and improve the electrical contacts before transferred to a Lakeshore Cryogenics vacuum probe station for electrical transport measurements. Additional current annealing was carried out on some devices to further remove adsorbed impurities prior to transport measurements in high vacuum (~ $10^{-6}$ torr).

To further elucidate the nature of the remaining disorder in the GNRs, we fabricated an additional top-gate with solid polymer electrolyte consisting of lithium perchlorate ($LiClO_4$) and poly(ethylene oxide) (PEO) in the 1:8 weigh ratio. The top-gate electrodes were simultaneously patterned on the substrate along with the drain and source electrodes. Electrical properties of the devices were measured by a semiconductor parameter analyzer (Keithley 4200) in vacuum ($1\times10^{-6}$ torr).

**Results and discussion**

To globally characterize the extent of exfoliation and the quality of the solution exfoliated graphene, we deposited the graphene on a Si substrate and performed Raman Spectroscopy measurements at several randomly selected locations. Nearly identical Raman spectra were obtained at all locations. Figure 1 shows a representative Raman spectrum exhibiting three bands: the D band ( ~ 1354 cm$^{-1}$), G band (~ 1582 cm$^{-1}$) and 2D band (~ 2712 cm$^{-1}$). For comparison, a Raman spectrum obtained on the starting graphite powder is also included in Figure 1. The shape of the 2D band for the solution exfoliated graphene clearly differs from that for the graphite powder, indicating that the majority of the exfoliated graphene flakes consist of few layers (< 5 layers). In addition, the D-to-G-band intensity ratio ($I_D/I_G$) for the solution exfoliated graphene (~0.5) is significantly higher than that in the starting graphite powder (~ 0). The presence of a D peak is usually attributed to edges and/or topological defects in the basal plane; and the $I_D/I_G$ increases with the overall structural disorder. Since the average size of the solution-exfoliated graphene flakes (typically hundreds of nanometers as discussed below in detail) is much smaller than the laser spot size ( ~ a few microns), the increased $I_D/I_G$ in the exfoliated graphene in comparison with the starting graphite can be largely attributed to the edges of the solution produced graphene [10]. On the other hand, the $I_D/I_G$ in our solution exfoliated graphene is still substantially smaller than that in chemically reduced graphene [16], suggesting that our solution produced graphene flakes contain lower disorder than in reduced graphene.

Next, AFM was used to characterize a large number of the solution exfoliated graphene samples deposited on a Si substrate. In addition to irregular-shaped graphene flakes, narrow ribbons of single-layer or few-layer graphene with relatively high aspect ratios (sub-10 nm to tens of nm wide and hundreds of nm to 1 μm long) were also routinely observed. Figure 2(a) and (c) show representative AFM images of a substrate surface deposited with graphene sheets and nanoribbons. Figure 2(b) and (d) are zoomed images of the two GNRs from the marked areas in figure 2(a) and (c), respectively. Their widths are 6~8 nm and 23~26 nm, respectively [14]. From their line scans [figure 2(e) and (f)], we estimate that the GNR in figure 2(b) is ~1 nm thick; and the GNR in figure 2(d) is ~ 2 nm thick. Based on the previously reported AFM results of graphene sheets and nanoribbons [2, 15], they are likely to contain one and few monolayers (3-4 layers), respectively. A few GNRs with widths less than 100 nm were observed in each randomly selected 10 μm × 10 μm area on the Si/SiO$_2$ substrates soaked in the GNR solution for 1 hr. Figure 3(a) and (b) show the topographic height and length distributions of over 100 GNRs with widths less than 100 nm as characterized by AFM. The minimum topographic height is 0.8 nm, and the heights of most GNRs fall into the range between 1 nm to 3 nm, suggesting that the GNRs are mostly few layers and consistent with the Raman spectroscopy data. The lengths for most of the GRNs range from 300 nm to 1 um, which allows us to fabricate FET devices on individual GNRs.

We emphasize that the use of PmPV polymer and humidity are two factors critical to producing GNRs with high aspect ratios. We were unable to produce GNRs without PmPV under the otherwise nominally identical conditions, consistent with the finding of Li *et al* in their study of GNRs produced from expandable graphite [15]. Similar to polyvinylpyeeolidone (PVP), which has been shown to help decrease the possibility of cutting long GNRs into shorter pieces [17], we suggest that the PmPV conjugated polymer known to non-covalently attach to graphene not only stabilizes the exfoliated graphene in solution but also reinforces the structural integrity of GNRs in the solutions, protecting them from breaking down to smaller particle-like structures. Furthermore, we have found that the yield of graphene ribbons or sheets significantly decreases when samples were prepared in dry laboratory air (relative humidity ≤ 15%) using dry graphite powder (exposed to only dry nitrogen or dry laboratory air) and anhydrous DEC solutions of PmPV. Increasing sonication time alone does not substantially increase the yield of thin graphene sheets and nanoribbons. Instead, excessive sonication (>2hr) only breaks graphite flakes into smaller particulates (tens of nanometers in lateral dimensions as well as in thickness). The yield of graphene ribbons and sheets noticeably increased when the graphite powder was exposed to relatively humid air (relative humidity > 25%) before mixing with DCE solutions of PmPV. While more detailed investigation of parameters affecting the exfoliation efficiency is needed, we tentatively attribute our experimental observation to water-molecule induced reduction of friction between adjacent graphene layers in graphite [18], which in turn facilitates more efficient exfoliate of graphene through sonication. A likely scenario is that the dangling bonds at the edge sites of graphite need to be saturated by molecules such as water molecules in order to maintain the low-friction behaviors [19, 20].

To characterize the electrical transport properties of individual GNRs, we have fabricated field-effect transistor devices consisting of individual GNRs. Figure 4 shows the electrical transport characteristics of a typical GNR device (L ~ 280 nm, W ~ 33 nm, and d~2 nm corresponding to about 3 layers). The AFM image of the device is depicted in the inset. The low-bias current-voltage (I-V) characteristics of the device are linear at all measured gate voltages [see figure 4(a)], indicating near Ohmic electrical contacts. As shown in figure 4(b), the transfer characteristic of the GNR device was p-doped with a charge neutrality point (CNP) beyond + 80 V before current annealing, which can be partially attributed to the adsorption of air or water molecules, or PMMA residue [21, 22] . After current annealing, the GNR exhibits ambipolar behavior with the minimum conductance associated with the CNP shifting to $V_g$ ~ 18 V, indicating that the adsorbed charge impurities have been partially removed. In addition, the overall conductance decreases after current annealing, which can be attributed to the lower carrier density due to the reduced impurity doping upon the removal of adsorbed impurities. From the transfer characteristics, the field effect mobility can be estimated as:

$\mu = [\Delta G \times (L/W)]/(C_{bg}\Delta V_g)$     (1)

Here $G$ is the low-bias conductance of the sample[23]; $L$ and $W$ are the channel length and width, respectively; and $C_{bg}$ is the back-gate capacitance (estimated to be $\sim 6\times10^{-8}$ F/cm$^2$ based on the capacitance of GNR-FET devices with similar ribbon width [24]). The hole mobility (both before and after current annealing) is ~20 cm$^2$/V s, and the electron mobility (after current annealing) is about ~ 5 cm$^2$/V. Similar mobility values are observed in three other few layer GNRs with the ribbon width ranging from 20 nm to 100 nm. These mobility values are over an order of magnitude higher than those reported for chemically reduced graphene oxide[25], suggesting lower disorder in our GNRs than in reduced graphene. However, they are still noticeably lower than the mobility values of GNRs produced from some other methods [26-28], suggesting that our GNRs still contain a substantial amount of disorder.

Carrier mobility in GNRs is largely determined by the phonon scattering, charged impurities on the GNR surfaces and in the SiO$_2$ substrate, structural defects in the basal plane, and edge-disorder [29]. Since our GNRs are formed by the tearing effect of bursting ultrasonic hot gas bubbles, we expect them to have smoother edges and lower edge disorder than lithographically defined GNRs [15, 27]. To further elucidate the nature of the remaining disorder in the GNRs, we have measured the temperature dependence of the transfer characteristics on some devices. Subsequently, we fabricated an additional polymer-electrolyte top-gate and measured their electrical transport characteristics under the top-gate configuration. A optical micrograph of a representative polymer-electrolyte top-gated GNR device is shown the inset of figure 5(b). Substantial performance improvement has been previously demonstrated in carbon nanotube FETs in the polymer-electrolyte top-gate configuration than in the back-gate configuration due to the enhanced gate-channel coupling [30]. Figure 5(a) shows the comparison of the transfer characteristics of another GNR (~ 200 nm long, ~ 50 nm wide and ~ 1.5 nm thick corresponding to 2 layers) operating in the back-gate configuration without polymer electrolyte and operating with a polymer electrolyte top-gate. We note that, without the polymer electrolyte, i) the GNR was p-doped with a CNP beyond the measured back-gate voltage range; and ii) the transfer characteristics of the device are nearly temperature independence for 4.3 < T < 297 K, in sharp contrast with the over 3 orders of magnitude conductance decrease in chemically reduced graphene upon cooling from room temperature to 4 K [16]. The nearly temperature independent transfer characteristics suggest that our solution exfoliated GNRs contain much lower structural disorder than chemically reduced graphene. The low-temperature on-off ratio of our GNRs is much smaller than that of typical lithographically-defined GNRs with comparable widths[31-33]. The high on-off ratio observed at low temperatures in the latter can be attributed to the opening of a transport gap near the CNP due the combined effects of a small confinement gap and disorder-induced potential fluctuations [33]. However, our GNRs are strongly p-doped and away from the CNP for the entire measured back-gate voltage range. The conductance versus polymer-electrolyte top-gate of the same device exhibits highly symmetric ambipolar behaviors. Moreover, the field effect hole-motility increases from ~ 11 cm$^2$/ V.s in the back gate configuration (prior to adding the polymer electrolyte ) to ~ 120 cm$^2$/V.s in the polymer electrolyte top gate configuration (the top-gate capacitance used to derive the field effect mobility is estimated to be ~ 1μF/cm$^2$ [34]; and the leak current remains below 500 pA within the top-gate voltage range examined).The highly symmetric transfer characteristic of the GNRs measured in the polymer-electrolyte top-gate configuration also enables a more accurate estimation of the carrier mobility by excluding the contact resistance contribution using the following model:

$$R_{total} = R_{contact} + R_{channel} = R_{contact} + \frac{L/W}{ne\mu}, \quad (1)$$

Here, $R_{contact}$ and $R_{channel}$ are the metal/GNR contact resistance and GNR channel resistance, respectively[35]; $L$ and $W$ are the channel length and width, respectively; $\mu$ is the carrier mobility, and the carrier concentration $n$, can in turn be determined by the expression,

$$n = \sqrt{n_o^2 + C_{tg}(V_{tg} - V_{CNP})^2}, \quad (2)$$

Here, $n_o$ is the residual carrier concentration at the maximum resistance, $C_{tg}$ is the polymer electrolyte top-gate capacitance (~ 1×10$^{-6}$ F/cm$^2$)[34], and $V_{CNP}$ is the gate voltage at the CNP [35, 36]. Besides the contact resistance, scattering mechanisms such as point defects and phonons may also contribute to the gate-independent resistance [37, 38]. However, these scattering mechanisms become significant only in high mobility graphene samples; and are expected to give very minor contributions to the gate-independent resistance in our GNRs with relatively low mobility (see detailed discussions below). As shown in Figure 5(b), this model fits our experimental data reasonably well, yielding a mobility value of 180 $cm^2/V$ in reasonable agreement with the field effect mobility (~120 $cm^2/V$). This mobility value is also comparable to the mobility of lithographically and chemically derived GNRs of similar widths [27, 32]. The slightly larger mobility value from the model fitting compared to the field effect mobility is largely due to the exclusion of the contact resistance. Although this model assumes a gate-independent contact resistance, we believe this is a reasonable assumption for our devices given the nearly ohmic contact and reasonably good fit of the data to the model, which is also consistent with the findings of Russo *et al.*[39].

On the one hand, the polymer electrolyte adds extra charged impurities to the GNRs, which is expected to reduce the mobility in the GNRs [40]. On the other hand, counter-ions in the polymer electrolyte accumulate on the GNR to neutralize the effects of charged impurities [41]. Two orders of magnitude increase of mobility has been previously observed in graphene FET devices immersed in ionic solutions, which was attributed to the ionic screening of charged-impurity scattering in graphene [41]. The order of magnitude mobility increase in our GNR devices upon the application of a polymer-electrolyte top-gate along with the nearly linear gate-dependence of the conductance indicates that the ionic-screening effect dominates in our samples. This in turn suggests that the charge transport in our GNRs is largely limited by charged impurity scattering. The screening-induced mobility-enhancement was also observed in the back-gate configuration (with polymer electrolyte) although at a lesser degree. The reduced screening effect in the back-gate configuration may be due to the electrostatic shielding of the top-layer in few-layer (2-3 layer) GNRs, while the electrolyte top-gate directly tunes the carrier density in the top-layer [42]. A likely source of charged impurities is the metallic impurities in the starting graphite powder. Charged impurities could also be introduced in the GNR synthesis and device fabrication processes. It is also worth noting that ion-screening is more effective than current annealing in reducing the charged impurity scattering in our substrate supported GNRs with initially relatively low mobility. While current annealing only partially removes impurities on the top surface of the GNRs, the counter ions in the polymer electrolyte are able to neutralize charged impurities not only on the top surface but also trapped beneath the ribbon. Due to the freezing of the ions in the polymer electrolyte at low temperatures, we were unable to measure the temperature dependence of the transport characteristics in the top-gate configuration.

In summary, we have developed a simple solution method to produce GNRs from graphite powder without any prior chemical treatment of the graphite. Single layer and few layer GNRs as narrow as sub-10 nm were observed by AFM. The main advantages of this method include 1) simplicity in production; and 2) free of covalent functionalization of GNRs. To characterize the electrical properties of the GNRs, we fabricated FET devices consisting of individual GNRs using electron beam lithography followed by the deposition of Cr/Au. The transfer characteristics of the devices show p-doping behavior, and the mobility of the devices is estimated to be at order of 10 $cm^2$/V s. Transport measurement of the same devices in the polymer-electrolyte top-gate configuration shows an order of magnitude mobility increase, which can be attributed to ionic screening of charged impurity scattering. We expect that higher mobility is achievable in these GNRs by further reducing the disorder induced by charged impurities.

Acknowledgement

Z.Z acknowledges the support of the Wayne State University Research funds. Part of this research was conducted at the Center for Nanophase Materials Sciences under project # CNMS2009-04

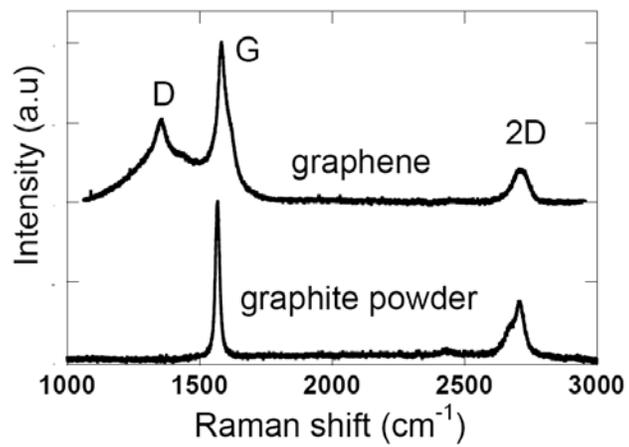

Fig. 1. Raman Spectra of solution exfoliated graphene and starting graphite powder. The intensity in each spectrum is normalized to its G-band intensity.

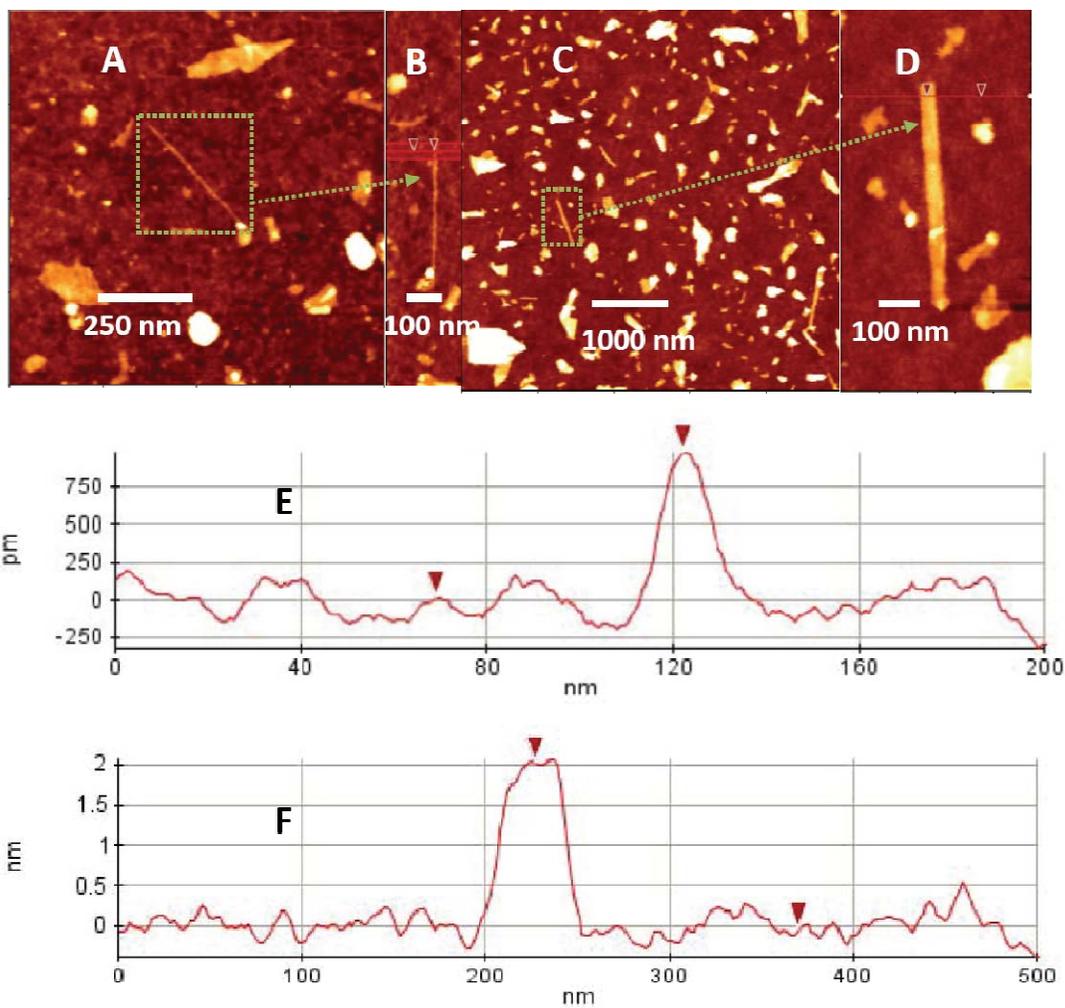

Fig. 2. GNRs produced from sonicating graphite power in DCE solution of PmPV. (a - d) AFM images of GNRs deposited on SiO$_2$ surface. (b and d) Zoomed images of the marked areas in (a) and (c), respectively: the width of the GNR in (b) is 6 ~8 nm; the width of the GNR in (d) is 23 ~ 26 nm. (e and f) Line profiles of the two GNRs in (b) and (d), respectively.

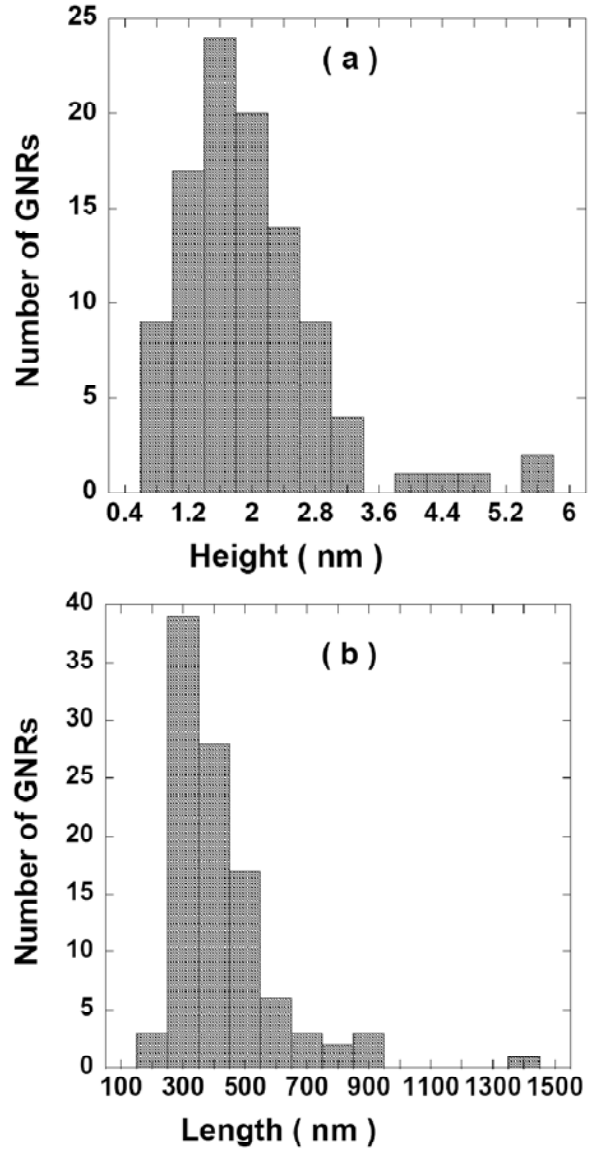

Fig. 3. Histograms of topographic heights (a) and lengths (b) of over 100 GNRs imaged by AFM on $SiO_2$ surface: graphene samples narrower than 100 nm with a length to width ratios large than 3 are included. GNRs with length- to-width ratio > 10 are routinely observed.

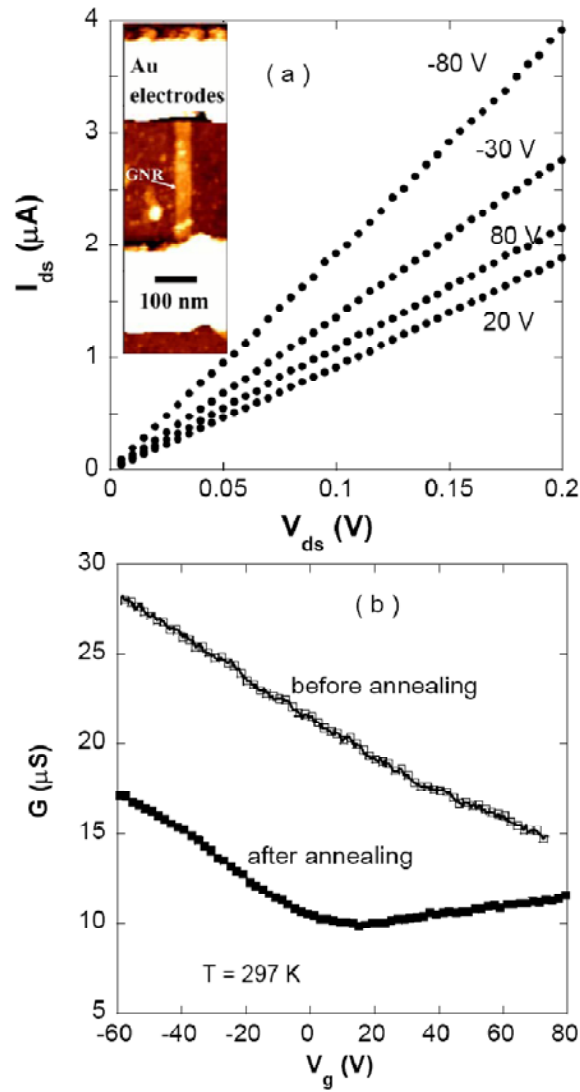

Fig. 4. Electrical properties of a GNR-FET device. (a) Drain-source current versus drain-source voltage measured at various back-gate voltages. (b) Room Conductance versus back-gate voltage measured at 297 K before and after current annealing . Inset: AFM of the measured GNR-FET devices with two Au electrodes contacting an individual GNR.

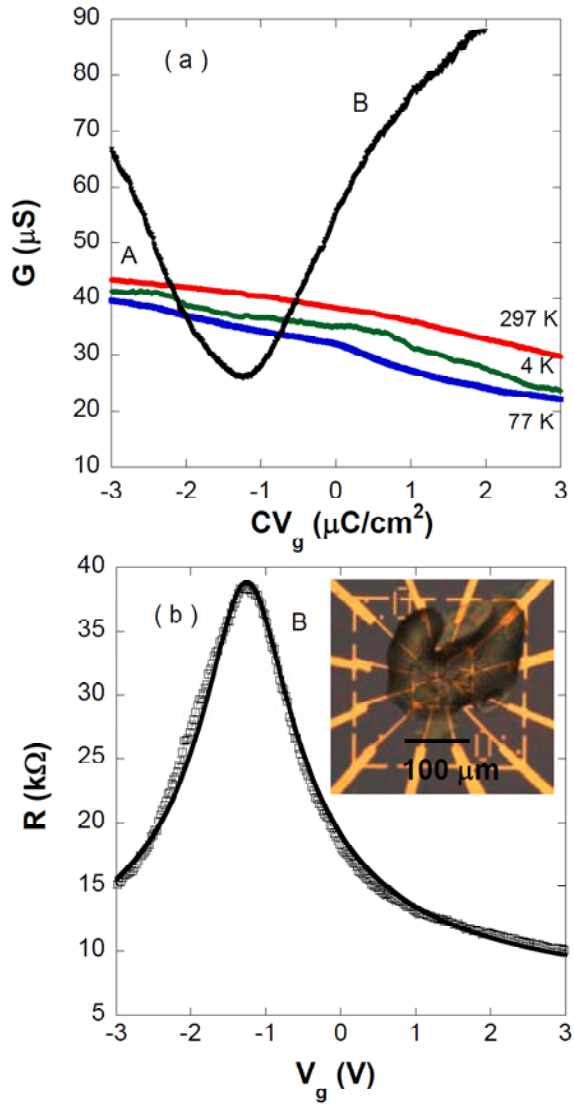

Fig. 5. (a) Comparison between back gating without the polymer electrolyte (A), and polymer top gating (B) on the same GNR-FET device. (b) Resistance versus polymer top-gate voltage for the same device. The solid lines in (b) are the model fitting. Inset: optical micrograph of the device with polymer electrolyte gate.